\documentstyle[12pt,aasms4]{article}

\lefthead{Jun}
\righthead{Pulsar Wind with Supernova Remnant}

\begin{document}

\title{Interaction of a Pulsar Wind with the Expanding Supernova Remnant}

\author{Byung-Il Jun\altaffilmark{1}}
\affil{Department of Astronomy, University of Minnesota,
    116 Church Street, S.E., Minneapolis, MN 55455}

\altaffiltext{1}{current address : Lawrence Livermore
National Laboratory, P.O. Box 808, L-630, Livermore, CA 94551, 
email : jun2@llnl.gov }

\begin{abstract}
Recent HST (Hubble Space Telescope)
observations of the Crab Nebula 
show filamentary structures that appear to originate from the
Rayleigh-Taylor (R-T) instability operating on the supernova ejecta
accelerated by the pulsar-driven wind.  In order to understand the
origin and formation of the filaments in the Crab Nebula, we
study the interaction of a pulsar wind with the uniformly expanding supernova
remnant by means of numerical simulation.
We derive the self-similar solution of this model for a
general power law density profile of supernova ejecta.

By performing two-dimensional numerical simulations, we find three
independent instabilities in the interaction region between the pulsar
wind and the expanding supernova remnant.
The first weak instability occurs in the very beginning, and is caused
by the impulsive acceleration of supernova ejecta by the pulsar
wind. The second instability occurs in the post-shock flow (shock wave
driven by pulsar bubble) during the intermediate stage. 
This second instability develops briefly while the
gradients of density and pressure are of opposite signs (satisfying
the criterion of the R-T instability). 
The third and most important instability develops
as the shock driven by the pulsar bubble becomes accelerated ($r \propto
t^{6/5}$). This is the strongest instability and produces pronounced 
filamentary structures that resemble the observed filaments in the
Crab Nebula.   Our numerical simulations can reproduce important
observational features of the Crab Nebula.  The high density heads in
the R-T finger tips are produced because of the compressibility of
the gas.  The density of these heads is found to be about 10 times
higher than other regions in the fingers.
The mass contained in
the R-T fingers is found to be $60 \% - 75 \%$ of the total shocked
mass and the kinetic energy within the R-T fingers is $55 \% - 72 \%$
of the total kinetic energy of the shocked flow.   The R-T fingers are
found to accelerate with a slower rate than the shock front, which is
consistent with the observations.  By comparing our simulations and the
observations, we infer that the some finger-like filaments (region F or
G in Hester's observation) started to develop about 657 years ago.

\end{abstract}



\section{Introduction}

Pulsars release their rotational energy in the forms of a relativistic
wind and electromagnetic waves (Rees and Gunn 1974) and this energy
generates the nebula that expands into the surrounding supernova remnant.
In the Crab Nebula, this spin-down power of pulsar is sufficient to support the
synchrotron radiation of the nebula, although the complete model to
link the pulsar to the nebula is yet to be given.  The pulsar wind is also
expected to interact with the expanding supernova remnant and produce
some interesting observational results.

The study of the interaction between the pulsar wind and
the enveloping supernova ejecta has been motivated since the recent
observation of the Crab Nebula by the Hubble Space Telescope (HST) 
revealed the detailed morphology and ionization structures of the
helium-rich filaments. 
The complex emission-line filaments would appear to 
originate from the Rayleigh-Taylor (R-T) instability operating on the
ejecta accelerated by the pulsar wind (Hester et al. 1996).    These
finger-like structures seem to grow inward (toward the center of the
remnant) and to be connected to each other at their origins by a faint,
thin tangential structure (the ``skin'').  Some of the long
finger-like structures terminate in dense ``head'' regions.
The interface between the synchrotron nebula and an ejecta shell has
been thought to be R-T unstable (Chevalier and Gull 1975; Chevalier
1977) since motions 
of the filaments (Trimble 1968) and the synchrotron nebula itself
(Bietenholtz et al. 1991) revealed their outward post-explosion 
acceleration.  The magnetic field in the Crab
has been inferred to be strong ( a few hundred $\mu G$) and seems to play
an important role in the formation of the filaments.  
Polarimetric VLA observations of the Crab
show a radially aligned magnetic field orientation (Bietenholz and
Kronberg 1990), although the rotating pulsar is likely to produce
toroidal (tangential) magnetic fields (Rees and Gunn 1974).  
The radial magnetic field can be produced as R-T fingers
stretch the existing field lines.

In the Crab Nebula, the observed mass and kinetic energy are much
smaller than expected from the typical supernova (see review by
Davidson and Fesen 1985). 
Several authors have proposed that the Crab formed from a normal Type-II
supernova remnant and that an outer fast-moving shell contains most of mass
and kinetic energy (Chevalier 1977, Kennel and Coroniti 1984), despite
the absence of convincing detection of the surrounding shell 
(e.g. Frail et al. 1995).  In this model for the Crab, the
pulsar wind pushes into the uniformly expanding supernova
ejecta. Therefore, the model includes 
four shocks from the outside inward : a supernova blast wave, a reverse shock
moving into the supernova ejecta, an outward facing shock wave driven
by expanding 
pulsar bubble, and a wind termination shock (Fig.1).   Note that the
forward shock front driven by the pulsar bubble is actually disturbed
by the instability (see Fig.5).    
The signature of the shock front driven by the pulsar bubble
has been found near the boundary of the Crab Nebula recently by
Sankrit and Hester (1997) although the existence
of the shock is debated, because the steepening of the radio spectral
index near  
the outer boundary of the Crab Nebula has not been found (Frail et
al. 1995; Bietenholtz et al. 1997).   Sankrit and Hester
find that the shock should be expanding with a velocity about $v = 150
km/sec$ into the freely-expanding supernova ejecta. 

The normal Type-II supernova remnant
model is particularly attractive because there is no need for
a peculiar low-energy supernova event and it can be applied to other
pulsar-powered remnants.
Also, the mechanism for the observed
acceleration is naturally explained, because the pulsar's wind shock
expands with the law $r \propto t^{6/5}$ in the self-similar stage if the
pulsar's luminosity is assumed to be constant (Chevalier 1977).  
In fact, the low-energy event model itself,
which is considered because of the discouraging observational results
on the absence of fast moving shell in the Crab Nebula,
has many problems to reconcile the observed features of the Crab
Nebula.  First, even low-energy supernova should produce an outer shock since
the supersonically moving gas generates a shock ahead of it.  This
shock should have a current expanding velocity, about $v_{shock} \sim
1400 km/sec$.  Therefore, the fact that no steepening of the spectral
index near the boundary of the Crab Nebula has been found cannot
support the low-energy event model.    Actually, this low energy event
model can generate a stronger shock than that of normal type-II
supernova model.  Second, the acceleration of both the line-emitting
filaments and synchrotron nebula and the well-resolved Rayleigh-Taylor
finger-like filaments pointing inward have not been explained by
the low-energy event model.
Currently, the only mechanism that can accelerate both the
synchrotron nebula and the filaments is the pulsar wind blowing
into the freely-expanding supernova ejecta around the Crab Nebula.

In this paper, we will present the results of our numerical
investigation of a Type-II supernova remnant model on the origin of
the filamentary structures in the Crab Nebula.   In \S 2, we study the
self-similar solution for our model.  In \S 3, our numerical
methods and initial condition for the simulation is described.  \S 4
presents our numerical results on the hydrodynamic evolution and fluid
instabilities of the interaction region.  In \S 5, we will discuss the
evolution of Rayleigh-Taylor fingers in terms of mass and kinetic
energy.  We discuss the related issues to the Crab Nebula further in \S
6 and summarize our main conclusions in \S 7.

\section{Self-Similar Solution}

  Chevalier(1977; 1984) has studied the self-similar solution for
the current model of the pulsar bubble expanding into a uniformly moving
medium of constant density.  In order to understand the dynamics of
the system and our numerical results, we generalize Chevalier's
self-similar solution for a power law density medium.    We only
consider the region between the contact discontinuity and the forward
shock driven by the pulsar bubble.
Using dimensional analysis, the expansion law can 
be obtained readily, $r \propto t^{6-n-l \over 5-n}$ where $n$ and $l$ are the
power law indices for the moving ejecta density ($\rho \propto r^{-n}$)
and the pulsar luminosity ($L \propto t^{-l}$).  For a constant pulsar
luminosity and uniform ejecta density, the bubble expands with the law,
$r \propto t^{1.2}$; that is accelerating.  It should be noted that
the solution is only applicable for $n < 3$ because of the finite mass
requirement ($M \propto {t^{n-3} r^{3-n} \over (3-n)}$)
as pointed out by Chevalier (1992).  Also, the bubble must
expand at least with constant velocity (no deceleration is allowed)
because the surrouding medium is freely expanding.  In
actual evolution, the bubble can decelerate at early stages while the
bubble velocity is much higher than the ejecta velocity, because the
medium motion can be negligible (see one-dimensional numerical result in
section 4).  However, as the bubble expansion approaches the
self-similar stage, the effect of the moving medium is not negligible and
the bubble expansion cannot decelerate.  Therefore, the self-similar
solution is only applicable for $l \leq 1$.

 In order to derive a self-similar solution for general power law
density profile of supernova ejecta, we can define the similarity
variables 
\begin{equation}
\zeta = {r \over At^a}
\end{equation}
\begin{equation}
\rho = t^{b-na}D(\zeta)
\end{equation}
\begin{equation}
v = aAt^{a-1}V(\zeta)
\end{equation}
\begin{equation}
p = a^2A^2t^{b-na+2a -2}P(\zeta)
\end{equation}
where $A$ is constant, $a = {6-n-l \over 5-n}$, and $b$ is an
expansion factor defined as $b=n-3$.  The expansion factor $b$ is 0 if the
surrouding medium is at rest.  The effect of a constant expansion of
the supernova ejecta is to change the density with
the law $\rho \propto t^{n-3}$ at the same radius.

 The one-dimensional fluid equations for a spherical coordinate
system are 
\begin{equation}
{\partial \rho \over \partial t} + v {\partial \rho \over \partial r} +
\rho {\partial v \over \partial r} + {2 \rho v \over r} = 0
\end{equation}
\begin{equation}
{\partial v \over \partial t} + v {\partial v \over \partial r} + {1
\over \rho}{\partial p \over \partial r} = 0
\end{equation}
\begin{equation}
{\partial p \over \partial t} + v {\partial p \over \partial r} +
\gamma p {\partial v \over \partial r} + {2 \gamma p v \over r} = 0
\end{equation}
where $\gamma$ is the adiabatic index.

Using the similarity variables, we can transform the fluid equations
into the ordinary differential equations
\begin{equation}
({b \over a} - n + {2 V \over \zeta})D + (V - \zeta){dD \over d
\zeta} + D{dV \over d \zeta} = 0
\end{equation}
\begin{equation}
{(a-1) \over a} V + (V - \zeta){dV \over d\zeta}+ {1 \over D}{dP \over
d\zeta} = 0
\end{equation}
\begin{equation}
( {b-2 \over a} -n +2 + {2 \gamma V \over \zeta})P + (V - \zeta){dP \over d
\zeta}+ \gamma P {dV \over d\zeta} = 0.
\end{equation}
These equations can be integrated from the shock front (driven by the
expanding bubble) using the following boundary conditions
\begin{equation}
D(1) = {\gamma + 1 \over \gamma -1}
\end{equation}
\begin{equation}
V(1) = 1 + {\gamma -1 \over \gamma +1}(c -1)
\end{equation}
\begin{equation}
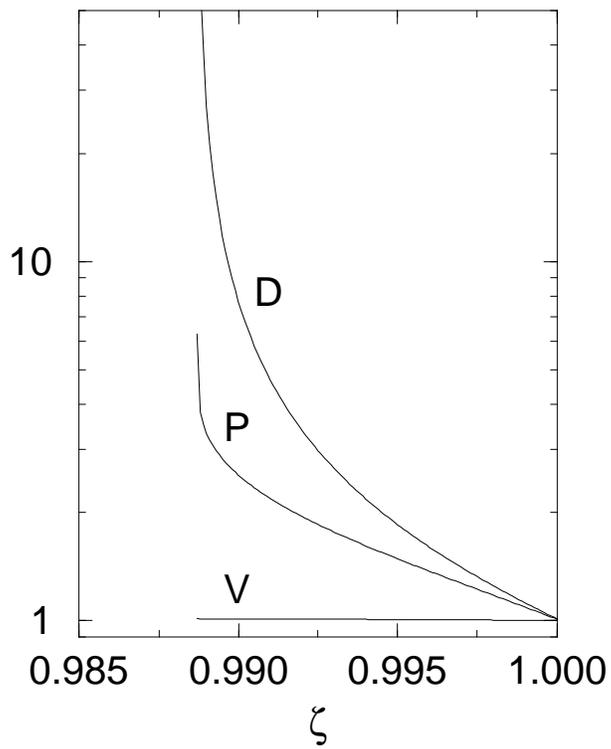
P(1) = {2 \over \gamma +1}( 1 - c)^2
\end{equation}
where $c$ is ${1 \over a}$ for a moving medium and 0 for a stationary
medium.

Self-similar solutions for several different cases are shown in Fig.2.
Adiabatic index is chosen to be 5/3 except for Fig.2d because the
shocked supernova ejecta region is considered to be
nonrelativistic while the pulsar wind is
relativistic.   Each plot shows density, gas pressure, and
velocity profiles in the region between the contact discontinuity and the shock
front.   In general, the flow within the shell (shocked supernova
ejecta) are found to be stable under the Rayleigh-Taylor instability criterion
in all cases since the gas pressure gradient has the same sign as the density
gradient.  But the contact discontinuity is Rayleigh-Taylor unstable
for $a >1$ because the shell is denser than the region inside of
the contact discontinuity.  Fig.2a is the case III in Chevalier
(1984).  The shell thickness is about $0.02 r_{shock}$.  Comparing
Figs. 2a, 2b, and 2c, the shell becomes thicker as $a$ increases as
also found by Chevalier (1984).   We find that the case in Fig.2b shows a
thicker shell than the case for $a=1.5, n=0$.   Fig.2d shows the
same case as Fig.2a except for $\gamma = 1.1$.  This case is
considered to see the effect of radiative cooling.  As expected, the
shell is found to be thinner in the case for $\gamma = 1.1$ than
$\gamma = 5/3$.

\section{Numerical Methods and Initial Condition}

Understanding the dynamical interaction between the
pulsar wind and the ejecta requires multidimensional numerical
simulations with high resolution because the development of the
instability is highly nonlinear.   High resolution simulations
are particulary necessary to resolve the thin shell and small regions
containing R-T fingers.   For this problem, we utilize the ZEUS-3D code, 
developed and tested at the National Center for
Supercomputing Application (Clarke \& Norman, 1994).  
ZEUS-3D is a three-dimensional Eulerian finite difference code for solving the 
ideal MHD equations.   The grid velocity is also allowed to
change every timestep so that the grid can follow the expanding
system.   This is necessary in modeling the Crab Nebula in
order to follow the expanding supernova remnant accurately and keep
adequately high resolution at the interaction region between the
pulsar wind and the supernova ejecta.     For the detailed numerical
algorithms of this code, readers are referred to Stone \& Norman (1992).

The model includes the pulsar wind, moving supernova ejecta, and
supernova shell confined by the supernova blast wave and reverse
shock.   Since we are interested in the interaction region
between the pulsar wind and the supernova ejecta, our simulation
excludes the supernova shell.   Readers are referred to Jun \& Norman
(1996a; 1996b) for detailed study of the dynamics of supernova shell.
The pulsar wind is generated by the
constant mechanical luminosity, $L = 1.0 \times 10^{40} ergs/sec$.
The luminosity of the pulsar is given by $L = 2.0 \times \pi r^2 \rho
v^3$ where $r$, $\rho$, $v$ are the initial radius of the wind, the
density of the wind gas, and the initial wind velocity, respectively.  This 
luminosity is much higher than the current luminosity of the
Crab pulsar, $L \sim 1.0 \times 10^{38} ergs/sec$, and is determined by
accounting for the observed rate of decrease in the luminosity.  
The initial radius of the wind is chosen to be $0.1 pc$ and the
density of the pulsar wind gas is assumed to be $\rho = 1.67 \times
10^{-24} gm/c.c$.  The initial wind velocity is determined from the
given luminosity accordingly.
The supernova ejecta is initialized to
be uniformly moving with $v = 500 km/s$ at $0.1 pc$, and the velocity
at larger radius is linearly proportional to the radius.  
The density profile of the supernova ejecta is usually modeled by a
constant density in inner core region and the power law density
profile ($\rho \propto r^{-n}$) in the outer region although the mass
ratio between two regions can change depending on the model.
According to Chevalier \& Fransson (1992), the forward shock driven by
the pulsar bubble is 
likely to be still in the inner core region during the constant
luminosity phase.   In our simulation, we assume the constant density
of supernova ejecta to be $\rho = 1.67 \times 10^{-22} gm/c.c$ over the
entire ejecta.     Note that the ejecta density
decreases with time because of ejecta expansion.  
We assume the entire fluid to be nonrelativistic gas with $\gamma = 5/3$, for
simplicity.    We defer a more general simulation with varying
luminosity and different density profile of the ejecta including
detailed physics such as magnetic field and cooling to a future
project.  Because of our idealized simple initial condition, we do not attempt
to compare the evolution with exact time scale to the Crab Nebula.
We focus our investigation on the development and the formation of the
filamentary structures by fluid instabilities.

The simulation is carried
out in a spherical coordinate system with the resolution 800x500 in the
$r-\phi$ plane.    The thin shell is resolved by about 16 grid cells
in the r-direction with this mesh size, because the thickness of the
thin shell is 
approximately 0.02 times the shock radius.  This resolution in the
shell remains same throughout the evolution because grids are expanding as
the forward shock moves out.      We update the outer
boundary condition in the r-direction to take into account the
condition of the moving supernova ejecta in every time step.
The boundary conditions in the $\phi-direction$ are assumed
to be periodic. The density field is perturbed with the amplitude of
$1 \%$ in the entire region.   Incoming flow from the outer r boundary
also includes the density perturbation in order to trigger the
instability within the postshock flow (see next section for the
instability).   We evolve a Lagrangian invariant passive quantity
(mass fraction function) to follow the contact discontinuity
between the supernova ejecta and the pulsar wind.   The mass fractions
are assigned to be 0.0 in the pulsar wind and 1.0 in the supernova ejecta.

\section{Hydrodynamic Evolution and Instabilities}

\subsection{One-Dimensional Results}

First, The global evolution of the flow is studied by one-dimensional
numerical simulations (Fig.3).   In the early stage, the shock wave
driven by the expanding pulsar bubble propagates much faster than
the moving supernova ejecta and the density of supernova ejecta can be
considered to be roughly constant (stationary medium stage). 
Figure 3a plots density, velocity,
and gas pressure, from top panel to bottom panel at t = 100 years.  
Three typical hydrodynamical
features; namely, a wind termination shock (W.T.), a contact
discontinuity (C.D.) 
between the wind bubble and the shocked ambient medium, and a
forward shock (F.S.) driven by pulsar bubble are clearly seen from
inside outward. These flow structures resemble the wind solution in a
stationary medium as described by Weaver et al. (1977).    
The shock expansion can be approximated as $r_{shock} \propto t^{3/5}$.
As the shock velocity decreases and becomes
comparable to the ejecta velocity, the ejecta can be no longer
considered as a stationary medium and the effect of decreasing density
of the ejecta becomes important (moving medium stage).  
Then the shock accelerates and enters the self-similar stage which is
described by $\rho_{ejecta} \propto t^{-3} $ and $ r_{shock} \propto
t^{6/5}$ (Chevalier 1977,1984).   This stage is illustrated in
Figure 3c.   Figure 3b shows the intermediate stage evolving from the
stationary medium stage to the moving medium stage.  One noticeable
feature is the thickness of the shell between the contact
discontinuity and the forward shock.  This shell thickness decreases
with time and becomes considerably thinner, about $0.02 \times r_{shock}$,
in the moving medium stage.  This is a feature predicted by the self-similar
solution.  The expansion rate of the contact discontinuity as a
function of time is shown in Figure 4.  Thick solid line represents
one-dimensional numerical result while thin solid lines show
constant expansion rates, $r \propto t^a, a = 2/5, 3/5, 1, 6/5$ from the
flattest to the steepest curve for comparison.    At the end of the
simulation, the expansion rate of the contact discontinuity reaches
about $a=1.189 $ which is close to the predicted value by the self-similar
solution for the constant luminosity and uniform ejecta density.
Recall that the expansion rate derived in the self-similar solution
for varying luminosity and power law density profile is $r \propto
t^{a}, a = {6 - n - l 
\over 5 - n}$, where $n$ is the power law index of the ejecta density
profile ($\rho \propto r^{-n}$)
and $l$ is the power law index for the pulsar luminosity ($L \propto
t^{-l}$). We find that the time scale for the system to reach
the self-similar stage is sensitive to the physical variables such as the
pulsar wind and background ejecta flow.  The system evolves faster for
the higher pulsar luminosity but slower for the higher supernova ejecta
density.  Also, the evolution takes shorter time for higher ejecta
velocity and higher density of pulsar wind gas for a given
luminosity.  The faster evolution for the higher density of pulsar
wind in a given pulsar luminosity is likely because the higher
momentum is allowed by choosing a higher density for a given luminosity
(recall that the pulsar mechanical luminosity is $L = 2.0 \times \pi
r^2 \rho v^3$).

\subsection{Hydrodynamic Instabilities : Two-Dimensional Numerical Simulation}

Numerically, we found three independent instabilities in the interaction region
between the pulsar wind and the supernova ejecta.
The first weak instability occurs in the very begining, and is caused
by the impulsive acceleration of dense ejecta by the low density
pulsar wind.   But this is a very weak instability and does not
grow significantly at the interface (see Fig.5a).
The second instability develops in the post-shock flow during the
intermediate stage (Fig.5b).
This second instability develops briefly while the
gradients of density and pressure are of opposite signs (satisfying
the criterion of the R-T instability).   This unstable flow develops
in the transition between the stationary medium stage (Weaver's
self-similar stage) and the moving medium stage (Chevalier's
self-similar stage).   In the stationary medium stage, both pressure
and density profiles in the post-shock region show positive gradients
(see Fig.3a).
On the other hand, in the moving medium stage, both pressure and
density in the post-shock region show negative gradients
(see Fig. 3c).  In each stage, the post-shock flow is stable.  
However, while the stationary medium stage evolves into the the moving
medium stage, the postshock pressure and density evolve by $P \propto
v^2 t^{-3}$ and $\rho \propto t^{-3}$, respectively.  
Therefore, the pressure decreases
more rapidly with time than the density if the power law index of
the shock expansion ($r \propto t^a$) is smaller than 1.0.  The different
evolutions of density and pressure result in the unstable flow
that shows a positive gradient in density profile but a negative
gradient in pressure profile (see Fig.3b).   
This unstable flow disappears when the shock expansion accelerates.
Small fingers in Figure 5b are the results of this second instability.
We note that the fingers are confined within the shell and
the outer shock front is not affected by this instability.
The first and second instabilities do not result in the formation of
significant long filaments but they play a role as seed perturbation
for next strong instability.

The third and most important instability develops in the contact
discontinuity between the pulsar wind and the shocked supernova ejecta
as the pulsar bubble becomes accelerated ($r \propto t^a, a > 1.0$).
This is the strongest instability and produces pronounced 
filamentary structures (Fig. 6).
The thin layer between the shock and the contact discontinuity is
severely distorted because of the instability (note that the shock front
is not affected by the first and second instabilities).  Figure 6
shows a number of thin fingers pointing toward the center.  At
later times, most of these thin fingers become unstable while
some fingers maintain their stability longer.  
Some unstable fingers are found to be detached from the shell.
The disruption of long thin fingers is
caused by a secondary Kelvin-Helmholtz (K-H) instability that develops
along the 
shear layer between the pulsar's low density nebula and the R-T finger.
The growth rate of the K-H instability decreases as the
relative density ratio between two gases increases.  
The formation of these stable long fingers is possible
because of the large density difference between the shocked thin layer and
the low density pulsar bubble.  Therefore,
fingers will be able to maintain their stable long structure for
longer evolution if the density in the pulsar nebula is lower.
Actually, this is what we expect to happen in the Crab Nebula.
The stability of fingers should be also affected by the strong magnetic
field in the Crab Nebula (Hester et al. 1996).

As the R-T fingers grow, the mixing layer becomes thicker at later
times.   The normalized 
thickness of the mixing layer is expected to increase with time as
long as the shell acceleration goes on.
The angle-averaged density of two-dimensional numerical simulation at t =
4000 years is compared to the one-dimensional result in Figure 7.    The
shell becomes much thicker and the average density in the shell is
decreased as results of mixing in two-dimensional numerical simulation
compared to the one-dimensional result. 

What is most noticeable is the formation of
dense heads on the R-T fingers, which can be compared to regions F or G in the
HST image of Hester et al. (1996).  These dense finger heads are
attributed to the compressibility of the gas.  The density in dense
finger heads is found to be about 10 times higher than in other regions
of the fingers.  The density is expected to increase if the cooling
is important.
The general morphology of our numerical results is compared to the HST
observation of the filaments in the Crab Nebula (see Hester et
al. 1996) and found to resemble the observations.

\section{Mass and Acceleration of R-T fingers}

Figure 8 shows the mass distribution of the shocked supernova ejecta as
functions of radius (left column) and density (right column).    
The histograms in the left panels illustrate the mass distribution
contained within a shell bounded by two radii ($r_1, r_2$).  The mass is
normalized by total mass.   The radius is also normalized by the
outer boundary of the computational plane.  Recall that the outer
boundary in the r-direction expands with the shock velocity and the shock
front is distorted because of the strong instability.  The right panels 
show the normalized mass distribution as a function of the density
normalized by the unshocked ejecta density.  The mass distribution of
shocked supernova material extends  farther inside with an increasing
evolution time as seen from top panel (t=1000 years) to bottom panel
(t=4000 years) due to the growth of R-T fingers.  
Also, the higher density fingers are found to form at later
times ( compare the mass distributions as a function of density
at 1000 years and 2000 years or later).  This result confirms that the
high density at the tip of R-T finger is actually generated by the compressible
flow pouring down to the fingers from the shell rather than simple
drifting material from the shell.

The total mass in the shell changes with time as the supernova material
flows in from the outer shock and some mass drains away through the
formation of the R-T fingers.  Table 1 shows the relative amount of mass
and kinetic energy confined
within the shell and R-T fingers at a given time.   The mass is
normalized by the total shocked supernova mass.
The thickness of the shell approaches to about $0.02 \times r_{shock}$
in the self-similar stage ($r \propto r^{1.2}$) where $r_{shock}$ is 
the radius of outer shock. 
Our two-dimensional numerical solution shows that the
outer shock is severely distorted by the instability and the thickness
of thin shell varies as a function of azimuthal angle.  In order to
estimate the mass within the shell approximately, we first take the
shell thickness as $0.02 \times r_{shock}(\phi)$ where
$r_{shock}(\phi)$ is the shock radius at $\phi$.   As a comparison, we
also listed each mass in the shell and the R-T fingers by taking the
shell thickness as $0.03 \times r_{shock}$ in Table 1(b).
Our results show that the relative mass amount in the
shell decreases from t=1000 years to t=2000 years and then increases
later.  This is because the R-T fingers, due to the acceleration, 
are not fully grown and small R-T fingers still exist close to the
shell.   Besides, the shell thickness at t=1000 years is greater than
$0.02 \times r_{shock}$.  By looking at the trend from t=2000 years to
t=4000 years, we can see that the relative mass in the shell increases with
time while the relative mass in R-T fingers decreases.  This means
that incoming mass flux from the outer shock is larger than outgoing
mass flux through the formation of R-T fingers.  Once the long R-T
fingers are fully developed, the mass confined in R-T fingers is roughly
about $60 \% - 75 \%$ of total shocked mass and larger than contained
in the shell ($25 \% - 40 \%$).  Kinetic energy also shows a similar
tendency to the mass.   Considering kinetic energy change only between
t=2000 yrs and t= 4000 yrs, about $55 \% - 72 \%$ of total shocked
kinetic energy is in the R-T fingers and $28 \% - 45 \%$ is in the
shell.  Also, the kinetic energy in the shell increases with time
while the kinetic energy in the R-T fingers decreases with time.
The slightly higher percentage of kinetic energy in the shell than
mass can be explained because the flow velocity in the
shell is higher than the velocity of the R-T finger.

Figure 9 shows the power law indices of expansions of R-T fingers and
forward shock front for the different periods of time.   
The radius of R-T fingers is measured at the tip of the fingers.  This
point is detected by 
averaging the mass fraction function and taking the location where the
angle-averaged mass fraction is 0.01 (recall that the mass fraction in
the pulsar wind was 0.0 and the supernova ejecta material had mass
fraction 1.0 in the begining).  First of all, the expansion rates for
both R-T fingers and shock front increase with time.  After about
t=1000 years, both the R-T fingers and shock front start to accelerate,
but the shock front achieves a higher expansion rate than the R-T fingers.
The expansion of the R-T fingers reaches an asymptotic limit at about
a = 1.04 to 
1.05 ($r \propto t^a$).   This expansion rate is a little higher than the free
expansion.   In earlier stage (t=100 - 500 yrs), the expansion rates
of R-T fingers and shock front are close to each other because the R-T
instability is not fully developed.  Once the R-T fingers grow due to
the instability, they are decoupled from the shell and they experience
smaller acceleration than the thin shell.  On the other hand,
the shock front keeps accelerating until the expansion rate reaches
the self-similar value (a=1.2).  In our simulation, the power law
index for the shock front expansion reached about 1.157 during
t=3000 - 4000 yrs.   We expect that the power law index of the
expansion rate of the shock front will reach its self-similar value as
we evolve the system further.
This result tells us about the important history of R-T fingers.  
The supernova material is first accelerated impulsively by the pulsar
wind and decelerates until the expansion of thin shell between the
contact discontinuity and the forward shock starts to accelerate as
the system approaches the self-similar stage.  Once the thin shell
accelerates, the R-T instability develops and the R-T fingers
experience less acceleration by decoupling from the shell and falling
down along the effective gravity vector.
According to our numerical results, we can infer that the accelerated
filaments in the Crab nebula should have been 
accelerated more efficiently right before the R-T fingers were not
completely decoupled from the thin shell.   We will discuss this
issue further in the next section.

\section{Application to the Crab Nebula and Discussion}

  Our simulation is simple in terms of relevant physics and initial
condition.   For example, we have not considered the effect of
magnetic field and radiative cooling.  Strong magnetic field is known
to affect the stability and growth of R-T fingers (see Jun,
Norman, and Stone 1995).  We expect that inclusion of magnetic field
in our numerical simulation may generate more stable long fingers since
the secondary K-H instability can be suppressed by the magnetic field.
Also, tangential magnetic field may enhance coupling between the thin
shell and the R-T fingers and slow down the inward velocity of the
R-T finger.    As a result, the expansion power law index of the R-T
fingers may be increased.
Radiative cooling is expected to increase the density in
thin shell and R-T fingers and to affect the growth of R-T instability.
The cooling process may change the mass distribution of the shocked
supernova ejecta.   Therefore, the inclusion of magnetic field and
radiative cooling in the model is important in order to explain the
detailed structure of the filaments.
Besides, due to the lack of
good information about initial conditions such as the pulsar wind,
the exact quantitive comparison to the Crab
Nebula should not be made at the moment.  Nevertheless, some
qualititive results can 
be applied to understand the Crab Nebula better.

Bietenholz et al. (1991) found that the synchrotron component of
the Crab Nebula has accelerated since the supernova explosion and the
acceleration of the synchrotron component may be larger than that of
the emission-line filaments.  This observation is explained well by
our numerical simulation since the forward shock front and thin shell
accelerates with a higher rate than the R-T fingers.  From the table 2
in Bietenholz et al. (1991), we can derive the power law index of the
expansion for the outer edge of the synchrotron nebula because the
obeservations were carried out in two different years, 1982, 1987.
Using the relation for the expansion power law index, $a =
{log(r_1/r_2) \over log(t_1/t_2)}$ where $r_1$ and $r_2$ are the radii
at two different ages, $t_1$ and $t_2$, we calculate the expansion
power law index, $a = 1.245 \sim 1.328$.  This power law index is
somewhat larger than the predicted (1.2) from the self-similar solution for the
constant luminosity and uniform ejecta density.  The discrepancy could
have two possible origins.  First, the observed expansion parameter in
Bietenholz et al. (1991) is only an approximate value because of the
difficulty in comparing two radio images.  Also, the exact date of
the observations used can change the result.
Second, the density in the
supernova ejecta may be a decreasing function of the radius.  For
example, the self-similar solution predicts the power law index to be
$1.25$ for the power law index of the ejecta density of 1.0 and
constant pulsar luminosity.
In order to calculate the expansion power law index for the
emission-line filaments, we use Trimble's result of the convergence
date, t = 1140 yrs, using two sets of data in 1950 and 1966.
Therefore we can derive the expansion parameter, $e = 0.9806$ using
the relation, $t_e = \delta t /(1-e)$ (see Bietenholz et al. 1991).
And we calculate the the expansion power law index, $a = 1.107$.  This
value is also larger than our numerical result, $a=1.04 \sim
1.05$.   In general, the observed expansion power law indices of
synchrotron nebula and filaments are reasonably close to our numerical
simulation.  This is an encouraging result which supports the current
model for the Crab Nebula.  More accurate future observations of
the expansion rate of each component are highly desirable.

With two known expansion power law indices, we can derive when the
R-T instability due to the acceleration began to develop in the Crab
Nebula.  We assume that the expansion rate has been contant since the
development of the R-T instability.
We can now write a relation among the length of the R-T finger, an
initial time of the R-T instability development, and the expansion power
law indics ($a_f$ for the filaments and $a_s$ for the shock front) of
the R-T fingers.
\begin{equation}
{\delta r \over r_s} = 1 - (t /t_0)^{a_f-a_s}
\end{equation}
where $\delta r$ is the length of the R-T finger, $r_s$ is the radius of
the forward shock front, $t$ is the current age of the Crab Nebula,
and $t_0$ is the initial time of the R-T instability development.
Now let's use Hester's observation to obtain the length of the R-T
finger.   By choosing the regions F or G in the Figure 1 of Hester et
al. (1996), 
we take the length of the R-T finger as about 0.2 times the radius
of the forward shock front.   Here we also assume that the shock
front is located near the top of the R-T finger and we use the
expansion power law index for the outer edge of the Crab Nebula as the
expansion power law index for the shock front.   Then we
derive the initial age
of the R-T instability development at the regions F or G, $t_0$ =
266 years.     This means that the R-T fingers in region F or G began
growing 657 years ago.  This is an interesting result with regard to the
history of the line-emitting filaments.  The filaments in the Crab
Nebula started to form around the age of 266 years and the
acceleration of the filaments should had been strongest at that time.
We should note that this initial time for the R-T
instability is an approximate estimate because of our assumptions including
constant pulsar luminosity.  
Other smaller fingers in regions D
imply two different interpretations for the timescale of the formation.
First, those fingers may have started developing more recently compared
to the fingers in regions F or G if the shell in region D experiences
same expansion rate as region F or G.  Second, if the fingers in
region D have developed at the same time as regions F or G, then the
expansion rate, that is acceleration of the shell, must have been
slower in that region than region F or G.  This spherically asymmetric
expansion rate could be caused by the asymmetric pulsar wind.
Observational measurements of expansion rate in different azimuthal
regions is required to check these ideas.    One also needs to
remember that we have not considered the effect of magnetic fields in
controling the growth of the R-T fingers.
Our simulation cannot explain long large filamentary structures
extending continuously from the 'skin' to near the center.  We speculate that
these structures could be formed as the pulsar's wind sweeps the
supernova ejecta which contain large degree of inhomogeneity in density.

\section{Conclusions}

Our numerical simulation of the interacting pulsar bubble with expanding
supernova ejecta produces well-developed filamentary structures which
resemble the filaments in the Crab Nebula.     Three instabilities are
found to develop independently in different evolutionary stages.
The first weak instability develops near the contact discontinuity 
at early stage of the evolution due to the impulsive acceleration of
the dense ejecta by the pulsar wind.  The second instability develops
within the postshock flow region between the shock front and the
contact discontinuity.  The second instability is a time-transient
phenomenon before the system develops into the self-similar flow.
The third Rayleigh-Taylor
instability driven by the acceleration of the thin shell which
develops in the later stage is found to be the strongest and provide a
main mechanism for the origin of the filaments. 

As a result of the third R-T instability, the forward shock front is
severely distorted, while the first two instabilities could not disturb
the shock front.  A number of long R-T fingers are generated by the
instability and these fingers become unstable (Kelvin-Helmholtz
instability) due to the relative motion between the fingers and
background flow.  
However, the development of the K-H instability can be delayed by
a higher density ratio between two fluids and strong magnetic field.
The fingers produce dense heads at their tips and these
dense finger heads are attributed to the compressibility of the gas.
The density of these heads is about 10 times higher than other parts of
the fingers and expected to increase in the presence of the cooling
effect.
After the long R-T fingers are fully developed, the mass contained in
the R-T fingers is found to be roughly $60\% - 75 \%$ of the total shocked
mass.   Kinetic energy within the R-T fingers is about $55\% - 72\%$
of the total kinetic energy in the shocked flow.
The R-T fingers are found to accelerate with a slower rate than the
shock front as they are decoupled from the shell.  In our simulation,
the expansion power law index (a) for the tip of R-T fingers is about $1.04
\sim 1.05$ while the index for the shell is about 1.157.  These
results are close to the observed values in the Crab Nebula
(1.107 for the line-filaments and 1.245 $\sim$ 1.328 for the outer
edge of the synchrotron nebula).   The small difference between our
numerical results and the observations could be improved if the
expanding supernova ejecta has the density profile of $\rho \propto
r^{-1}$ or $r^{-2}$.  Considering the different expansion rates between
the outer edge of the synchrotron nebula and the line-emitting
filaments, we can infer that the several long fingers (region F or G
in Hester's observation) in the Crab Nebula 
started to grow about 657 years ago.

In summary, our model can explain important observational features
of the Crab Nebula. They are the accelerations of the filaments and
the synchrotron 
nebula,  formation of thin skin connecting the filaments,
formation of the filaments pointing toward the center, 
and high-density finger heads.


\acknowledgments

I am grateful to Roger Chevalier, Jeff Hester, Tom Jones, Mike Norman,
and Jim Stone for useful discussion and encouragement.    The work
reported here is supported by NSF grant AST-9318959 and by the
Minnesota Supercomputer Institute.

\clearpage

\figcaption[fig1.ps]{Schematic representation of the normal type-II
supernova remnant model for the simulation.  The figure shows the
pulsar wind blowing into the uniformly expanding supernova remnant.
The pulsar is located in the center.  The small dotted circle outside of
the pulsar is the wind termination shock.   This shock heats
up the wind gas and produces a hot bubble expanding into the supernova
remnant.    The contact discontinuity between the pulsar bubble and
the shocked supernova gas is shown to be Rayleigh-Taylor unstable.
The R-T fingers are connected to each other by a thin
shell confined by the forward shock.  Note that this forward
shock front is actually distorted by the development of the instability
in the numerical simulation (see Fig.6).
Outside the forward shock, there are expanding supernova ejecta,
reverse shock, and blast wave of the supernova. \label{fig1.ps}} 

\figcaption[fig2.ps]{Self-similar solutions for the model described in
Fig.1.  D, P, and V represent the density, gas pressure, and the
velocity, respectively.   Each plot shows the physical variables in
the  region between the contact discontinuity and the shock front
($\zeta = 1$). \label{fig2.ps}}

\figcaption[fig3.ps]{One-dimensional numerical simulation of the
normal type-II supernova remnant model.  Fig.3a shows the result at the
stationary medium stage.  From inside outward, the wind termination
shock (W.T.), contact discontinuity (C.D.), and forward shock (F.S.)
are seen at the designated location.    Each plot in the same column
represents the density, velocity, and gas pressure from top to bottom,
respectively.
Fig.3b and 3c are the results at the intermediate stage and the
moving medium stage, respectively (see the text for the
detail). \label{fig3.ps}} 

\figcaption[fig4.ps]{The figure plots the radius of the contact
discontinuity as a function of time.  The thick solid line represents
the one-dimensional numerical result while other thin solid lines show
constant expansion rate,$ r \propto t^a, a = 2/5, 3/5, 1, 6/5$ from the
lowest to the highest. \label{fig4.ps}}

\figcaption[fig5.ps]{The first image (left) shows the density of
two-dimensional numerical result at t=100 yrs.  From the center, each
circle corresponds to the contact discontinuity, the forward shock
driven by the pulsar bubble, and computational boundary.  The
computational boundary is round because the simulation is carried out
in the spherical geometry and remapped onto the cartesian grids.  The
wind termination shock is located inside the conatct discontinuity.
It is not visible in this image because of low density (see Fig.3).
The development of a very weak instability near the contact
discontinuity is seen.
The right image shows the result at t=600 yrs.  The forward shock is
seen near the outer computational boundary.   A number of small
fingers are developed in the postshock flow because of the R-T
instability in the flow.  Lighter color represents higher density.
\label{fig5.ps}} 

\figcaption[fig6.ps]{The density images of two-dimensional numerical
simulation at t=1000 (Fig.6a), 2000 (Fig.6b), 3000 (Fig.6c), 4000 (Fig.6d)
yrs.   Note that the forward shock front is always located near the outer
computational boundary because the grid is moving with the shock
velocity.   Lighter color represents higher density. \label{fig6.ps}}

\figcaption[fig7.ps]{The comparison between one-dimensional numerical
result and two-dimensional result at t=4000 yrs.  Two-dimensional
result is averaged along the angle-direction. \label{fig7.ps}}

\figcaption[fig8.ps]{Mass distribution of the shocked supernova ejecta
as a function of radius (left plots) and density (right plots).  Each
plot corresponds to the mass distribution at t=1000, 2000, 3000, and
4000 yrs from top to bottom. \label{fig8.ps}}

\figcaption[fig9.ps]{The expansion rate of the R-T fingers and shock
front as function of different period of time.  The figure plots power
law index of expansion, $a = {log(r_1/r_2) \over log(t_1/t_2)}$.
The power law indices for the shock front and the R-T fingers are
represented by circles and rectangles, respectively. \label{fig9.ps}}

\end{document}